\documentclass[letterpaper,pra,twocolumn,showpacs]{revtex4}

\usepackage{graphicx}
\usepackage{amsmath}
\usepackage{amssymb}

\begin{document}

\title{TEM$_{10}$ homodyne detection as an optimal small displacement and tilt measurements scheme}
\author{V.Delaubert$^{1,2}$, N.Treps$^{1}$, M.Lassen$^{2,3}$, C.C.Harb$^{2}$, C.Fabre$^{1}$, P.K.Lam$^{2}$ and H-A.Bachor$^{2}$}
\affiliation{$^{1}$Laboratoire Kastler Brossel, 4 place Jussieu,
case 74, Paris 75252 Cedex 05, France} \affiliation{$^{2}$ACQAO,
The Australian National University, Canberra ACT 0200, Australia}
\affiliation{$^{3}$Department of Physics, DTU, Building 309, 2800
Lyngby, Denmark }
%\normalsize{$^\ast$E-mail:  treps@spectro.jussieu.fr}

\date{\today}

\begin{abstract}
We report an experimental demonstration of optimal measurements of
small displacement and tilt of a Gaussian beam - two conjugate
variables - involving a homodyne detection with a TEM$_{10}$ local
oscillator. We verify that  the  standard split detection is only
64\% efficient. We also show a displacement measurement beyond the
quantum noise limit, using a squeezed vacuum TEM$_{10}$ mode
within the input beam.
\end{abstract}

\pacs{42.50.Dv; 42.30.-d; 42.50.Lc}
%42.30.-d Imaging and optical processing
%42.50.-p Quantum optics
%42.50.Dv Nonclassical field states
%42.50.Lc Quantum fluctuations, quantum noise, and quantum jumps

\maketitle

\section*{Introduction}

Measuring the transverse position of a laser beam seems to be a
very basic task. One could think that the best way to retrieve
such a simple information has been found years ago. However, we
have proven recently that the detection devices which are
traditionally used - split and quadrant detectors - are limited to
an efficiency of $64~\%$ \cite{Hsu-disp}. This is a question of
potentially great interest as they are used in many ultra
sensitive applications including optical tweezers, atomic force
microscopes, beam positioning for gravitational wave detectors and
satellite alignment \cite{Morrison, Chow, more}.

\begin{figure}[htbp]
\begin{center}
\includegraphics[width=8cm]{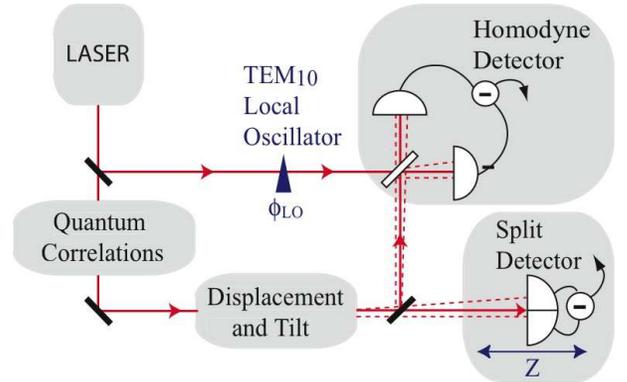}
\caption{{\bf Block diagram of our experimental setup. We compare
the performances of two measurement schemes - split detection and
homodyne detection with a TEM$_{10}$ local oscillator - to
retrieve displacement and tilt of a Gaussian beam. We show how to
modify the input beam using quantum correlations to perform
measurements beyond the Quantum Noise Limit.}} \label{intro}
\end{center}
\end{figure}

The purpose of this paper is to demonstrate the best possible
measurements of beam displacement and tilt, for a given beam power
and shape. Note that similar issues are addressed on the limits to
the measurement of beam rotation about its propagation axis in
reference \cite{Barnett}. We will first focus on the detection
device, and will therefore analyse experimentally a
split-detection and a homodyne detection scheme with a TEM$_{10}$
local oscillator (see Fig.~\ref{intro}), introduced theoretically
in reference \cite{Hsu-disp}. At the same time, we will present
measurements of the quantum conjugated variable of the transverse
displacement of the beam, the laser beam tilt \cite{Hsu-entang}.
Finally, we will focus on the laser beam itself, and show
measurements beyond the fundamental limit imposed by the photon
statistics of laser beams, using non classical beams. This
demonstration allows displacement and tilt measurements that were
masked or altered by quantum noise.

The paper is organized in the following way. We first give a brief
definition of displacement and tilt of a TEM$_{00}$ mode beam, and
introduce the notions of position and momentum of a Gaussian beam,
which are two conjugate transverse observables. In section
\ref{QNL}, we quantitatively discuss the QNL for displacement and
tilt measurements and show the improvement that can be achieved
with squeezed light. In section \ref{split}, we present how this
new set of quantum variables can be accessed with a split
detection, the conventional scheme used for beam displacement
measurements. The results obtained  provide a reference for a
homodyne detector with a TEM$_{10}$ local oscillator presented in
section \ref{homodyne}. In section \ref{SQZ}, we show how to
perform sub QNL measurements with both schemes and present
experimental homodyne detection results in this regime. In the
last section, a comparison between both schemes is presented, in
perfect agreement with theoretical predictions, and showing an
improvement with the homodyne detection that matches the predicted
detection efficiency of $64\%$ .
%

%%%%%%%%%%%%%%%%%%%%%%%%%%%%%%%%%%%%%
\section{Displacement and tilt of a Gaussian beam}

\begin{figure}[htbp]
\begin{center}
\includegraphics[width=8cm]{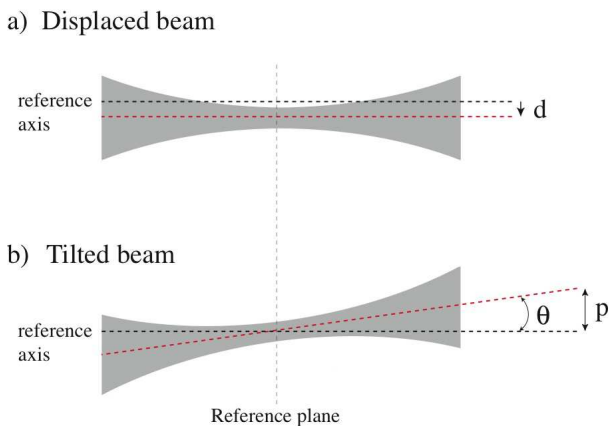}
\caption{{\bf Displacement and tilt definition. In the
$2$-dimensional case, the displacement $d$ corresponds to a
translation of the beam in the transverse direction. The tilt
$\theta$ corresponds to a rotation of the propagation axis. }}
\label{D_and_T}
\end{center}
\end{figure}

Displacement and tilt of a single-mode TEM$_{00}$ laser beam are
very intuitive notions, they refer to macroscopic properties of a
beam, as shown in figure (\ref{D_and_T}). We assume here that the
beam is constrained to one dimension, namely the figure plane of
the paper, considering that the non represented transverse
component is gaussian. A displacement corresponds to a translation
of the beam by a distance $d$ along the transverse direction,
whereas a tilt corresponds to a rotation of the propagation axis
by an angle $\theta$. The tilt of a laser beam is linked to the
transverse momentum of the beam, in the limit of small angles,
given by the following expression
\begin{eqnarray}
p=\frac{2\pi \sin{\theta}}{\lambda} \simeq \frac{2\pi
\theta}{\lambda},
\end{eqnarray}
where $\lambda$ is the optical wavelength. Note that displacement
and tilt are defined relative to a particular transverse reference
plane. For instance, in figure (\ref{D_and_T}), we have chosen the
beam waist plane as reference transverse plane.

In the case of small displacement and tilt, i.e. for $d \ll w_{0}$
and $\theta \ll \lambda/w_{0}$ - where $w_{0}$ is the beam waist
of the incident TEM$_{00}$ mode, we can Taylor expand the
displaced $E_d(x)$ and tilted $E_p(x)$ gaussian field to first
order, yielding \cite{Hsu-disp,Hsu-entang}
\begin{eqnarray}\label{field2}
E_d(x)\approx E(x) + d\cdot\frac{\partial E(x)}{\partial x} \\
E_p(x)\approx E(x) + i p \cdot x E(x).
\end{eqnarray}
The equations can be rewritten into
\begin{eqnarray}\label{field1}
E_{d,p}(x) = A_{0}\left[ u_{0}(x)+\left(\frac{d}{w_{0}} +i
\frac{w_{0} p }{2}\right)u_{1}(x)\right],
\end{eqnarray}
where $u_{n}(x)$ refers to the Hermite Gauss $TEM_{n0}$ mode
\cite{Siegman}. $A_{0}$ is the amplitude of the incident
TEM$_{00}$ mode and identifies with the one of the displaced and
tilted beam at first order in $d$ and $p$. Eq. (\ref{field1})
shows that the information of displacement and tilt of a
TEM$_{00}$ laser beam can be extracted by measuring the TEM$_{10}$
mode component of the field. Any displacement modulation is
transferred to the in-phase amplitude of the TEM$_{10}$ mode
relative to the "carrier" (TEM$_{00}$ mode), whereas any tilt
modulation is transferred to the TEM$_{10}$ component in
quadrature relative to the TEM$_{00}$ mode.

In order to give a quantum mechanical description of displacement
and tilt of a laser beam, we need to take into account the quantum
noise of all the optical modes of the beam, including the vacuum
modes. We can write the positive frequency part of the electric
field operator in terms of photon annihilation operators
$\hat{a}$. The field operator is then given in its more general
form by:
\begin{eqnarray}
\hat{\mathcal{E}}^{+}(x) = i \sqrt{\frac{\hbar \omega }{2
\epsilon_{0} c T}} \sum_{i=0}^{\infty}\hat{a}_{n}u_{n}(x),
\end{eqnarray}
where $\omega$ is the field frequency, $T$ is the integration
time, $u_{n}(x)$ are the transverse beam amplitude functions of
the TEM$_{n0}$ modes, and $\hat{a}_{n}$ are the corresponding
annihilation operators. $\hat{a}_{n}$ can be written in the form
of $\hat{a}_{n} = \langle\hat{a}_{n}\rangle + \delta\hat{a}_{n}$,
where $\langle\hat{a}_{n}\rangle$ describes the coherent amplitude
and $\delta\hat{a}_{n}$ is the quantum noise operator assumed
small in the linearization approximation.

For a mean photon number $N$, defined
by$|\langle\hat{a}_{0}\rangle|^{2}$ in the small displacement and
tilt regime, the quantum counterpart of equation (\ref{field1}),
assuming that only $u_{0}$ and $u_{1}$ are non-vacuum modes, is
\begin{eqnarray}\label{field}
\hat{\mathcal{E}}^{+}(x) &=& i \sqrt{\frac{\hbar \omega}{2
\epsilon_{0} c T}}\left[
\sqrt{N}\left(u_{0}(x)+\left(\frac{d}{w_{0}} +i \frac{w_{0} p
}{2}\right)u_{1}(x)\right)\right. \nonumber
\\
&& \left. + \sum_{i=0}^{\infty}\delta\hat{a}_{n}u_{n}(x)\right],
\end{eqnarray}
where we have introduced the mean value of position and momentum
quantum operators of a laser beam, $d=\langle\hat{x}\rangle$ and
$p=\langle\hat{p}\rangle$, respectively. These quantum operators
are given by
\begin{eqnarray}
\hat{x} &=& \frac{w_{0}}{2\sqrt{N}}\hat{X}^{+}_{a_{1}}\\
\hat{p} &=&\frac{i}{w_{0}\sqrt{N}}\hat{X}^{-}_{a_{1}},
\end{eqnarray}
where we see that position and momentum are linked to the
amplitude and phase quadrature of the TEM$_{10}$ mode component of
the field, respectively given by
\begin{eqnarray}
\hat{X}^{+}_{a_{1}}=\left(\hat{a}_{1}+\hat{a}_{1}^{\dagger}\right)\\
\hat{X}^{-}_{a_{1}}=i\left(\hat{a}_{1}-\hat{a}_{1}^{\dagger}\right).
\end{eqnarray}

Moreover, position and momentum are conjugate observables and
satisfy the following commutation relation \cite{Hsu-entang}
\begin{eqnarray}
\left[\hat{x},\hat{p}\right] = \frac{i}{N}.
\end{eqnarray}

In this section, we have defined displacement and tilt of a
Gaussian beam, given the general quantum description of such a
field, and shown that these former quantities were closely linked
to position and momentum, two quantum observables. Using this
interesting property, reference \cite{Hsu-entang} already proposed
a scheme for continuous variable spatial entanglement for bright
optical beams, involving two beams respectively squeezed in
position and momentum mixed on a $50:50$ beam-splitter.

%%%%%%%%%%%%%%%%%%%%%%%%%%%%%%%%%%%%%%%%%%%%%%%%
\section{Quantum Noise Limit for Displacement and Tilt Measurements}\label{QNL}

The use of classical resources (i.e. coherent laser beams) sets a
lower bound to detection performances, which is called the quantum
noise limit (QNL) and arises from the random time arrival of
photons on the detector. In the case of displacement measurement
of a laser beam, the transverse displacement $d_{QNL}$ of a
TEM$_{00}$ laser beam corresponding to a Signal to Noise Ratio
(SNR) of $1$, is found to be \cite{fouet,treps2}
\begin{eqnarray}
d_{QNL}=\frac{w_{0}}{2\sqrt{N}},
\end{eqnarray}
where $w_{0}$ is the waist of the beam, and $N$ its total number
of photons detected in the interval $T=1/RBW$, where $RBW$ is the
resolution bandwidth. Ideally, $T$ is maximized according to the
stability of the physical system. For instance in the case of bits
read-out in optical disc devices, $RBW$ roughly corresponds to the
scanning frequency. For a $100~\mu$m waist, $1~$mW of power at a
wavelength of $\lambda=1~\mu $m, with $RBW=100~$kHz, the quantum
noise limit is for instance given by $d_{QNL}=0.2~$nm. Note that
during test or characterization procedures, the precision can be
increased by averaging with the spectrum analyzer, for instance by
reducing the video bandwidth (VBW). The QNL effectively
corresponds to the minimum measurable displacement when $VBW=RBW$,
without averaging.

Similarly, the QNL for momentum measurements can be defined as
\begin{eqnarray}
p_{QNL}=\frac{1}{w_{0}\sqrt{N}}.
\end{eqnarray}
In the same conditions as the ones defined above, the QNL for
momentum measurement is $p_{QNL}=4.10^{-2}~$m$^{-1}$,
corresponding to a tilt angle of $\theta_{QNL}=7~n$rad.

In order to perform measurements beyond the QNL, i.e. for a given
$T$, we have shown in reference \cite{treps-multipixel} that
filling the appropriate transverse mode of the input field with
squeezed light is a necessary and sufficient condition. We call
this mode the noise-mode of detection \cite{Delaubert1}.

For example, using $3~$dB of squeezing in the appropriate
component of the beam for a displacement measurement leads to a
noise reduction of a factor 2. The SNR is quadratic in $d$ as the
signal corresponds to the intensity of the TEM$_{10}$ component of
the displaced field, and the new quantum limit is thus given by
$d_{SQZ}=d_{QNL}/\sqrt{2}$. It is important to note that, as
imposed by Heisenberg inequalities, the measurement of the
conjugated observable - the momentum in this case - is degraded.

%%%%%%%%%%%%%%%%%%%%%%%%%%%%%%%%

\section{Split detection}\label{split}

\subsection{Theory}

\begin{figure}[htbp]
\begin{center}
\includegraphics[width=8cm]{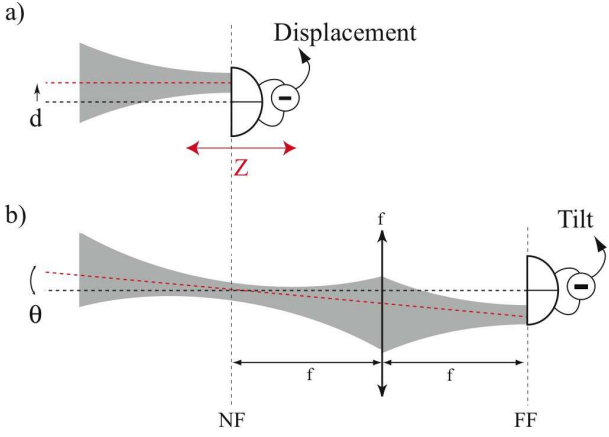}
\caption{{\bf Measuring displacement and tilt of a Gaussian beam
with a split detector. Taking a reference plane where displacement
and tilt components are needed, displacement can be measured
directly with a split detection in the near field (NF) of the
reference plane, whereas tilt can be accessed in its far field
(FF).}} \label{D_and_T_with_SD2}
\end{center}
\end{figure}

The conventional way to measure the displacement of a laser beam
is to use a split detector. As shown in figure
(\ref{D_and_T_with_SD2}a), the difference between the intensity on
each side of the split detector yields a photocurrent proportional
to the displacement. This technique is widely used notably for
beam alignments, particle tracking and atomic force microscopy.
Nevertheless, such a detection device only accesses the beam
position in the detector plane, and is totally insensitive to the
orientation of the propagation axis of the beam (i.e. tilt).
Consider the evolution of the field operator of Eq.~\ref{field}
under propagation along the $z$ axis, we get
\begin{eqnarray}\label{field_propag}
\hat{\mathcal{E}}^{+}(x,z) &=& i \sqrt{\frac{\hbar \omega}{2
\epsilon_{0} c T}}\left[ \sqrt{N}u_{0}(x,z)\right. \nonumber
\\ && \left. +\sqrt{N}\left(\frac{d}{w_{0}} +i \frac{w_{0}p }{2}\right)u_{1}(x,z)e^{i\phi_{G}(z)}\right.\nonumber
\\ && \left. + \sum_{i=0}^{\infty}\delta\hat{a}_{n}u_{n}(x,z)e^{i n
\phi_{G}(z)}\right],
\end{eqnarray}
where $u_{n}(x,z)$  is the Hermite Gauss $TEM_{n0}$ mode,
$\phi_{G}(z)$ is the Gouy phase shift, which equals
$arctan(z/z_{R})$, where $z_{R}$ is the Rayleigh range of the
beam. The displacement and tilt ratio varies along $z$ because of
the Gouy phase shift (i.e. diffraction), up to be perfectly
inverted in the far field where $\phi_{G}(\infty)=\pi/2$. This
Fourier Transform relation is a well known result in classical
optics, for which a displacement in the focal plane of a simple
lens is changed into an inclination relative to the propagation
axis. Therefore, if the exact amount of tilt and displacement is
needed in a particular transverse plane, for instance at $z=0$,
displacement can be measured in this plane (or in its near field),
whereas tilt can only be accessed in its far field, as presented
in figure (\ref{D_and_T_with_SD2}b).

The field presented in Eq.~\ref{field_propag} is detected via a
split detector whose position is varied along the $z$ axis. The
photocurrent is directly proportional to the difference of
intensity incident on the two halves of the detector
\begin{eqnarray}
\hat{I}_{-}(z)&=&\int_{0}^{\infty}\hat{\mathcal{E}}^{+}(x,z)\hat{\mathcal{E}}^{+\dagger}(x,z)dx\nonumber\\
&&-\int_{-\infty}^{0}\hat{\mathcal{E}}^{+}(x,z)\hat{\mathcal{E}}^{+\dagger}(x,z)dx,
\end{eqnarray}
and replacing $\hat{\mathcal{E}}^{+}(x,z)$ with the previous
expression yields, for very small displacement and tilt
\begin{eqnarray}\label{Iminus}
\hat{I}_{-}(z)&=&\frac{\hbar \omega}{2 \epsilon_{0} c
T}\left[2Nc_{1}\left(\frac{d}{w_{0}}\cos{\phi_{G}(z)}+\frac{w_{0}
p }{2}\sin{\phi_{G}(z)}\right)\right.\nonumber\\
&&\left.+\sqrt{N}\sum_{p=0}^{\infty}c_{2p+1}\delta
\hat{X}_{2p+1}^{- (2p+1)\phi_{G}(z)}\right],
\end{eqnarray}
where $ \delta \hat{X}_{n}^{\phi}=\delta
\hat{a}_{n}e^{-i\phi}+\delta \hat{a}_{n}^{\dagger}e^{i\phi}$
refers to the noise of the quadrature of the TEM$_{n,0}$ mode
defined by the angle $\phi$, and
\begin{eqnarray}
c_{n}&=&\int_{0}^{\infty}-\int_{-\infty}^{0}u_{n}(x)u_{0}(x)dx\nonumber \\
&=&\int_{-\infty}^{\infty}u_{n}(x)u_{f}(x)dx,
\end{eqnarray}
where $u_{f}$ is the flipped mode, which is a TEM$_{00}$ mode
whose transverse profile has a $\pi$ phase shift at the origin for
$z=0$ \cite{Delaubert-flipped}. Its decomposition in the
TEM$_{pq}$ basis during propagation is given by
\begin{eqnarray}
u_{f}(x,z)=\sum_{p=0}^{\infty}c_{2p+1}u_{2p+1}(x,z)e^{i(2p+1)\phi_{G}(z)},
\end{eqnarray}
and the fluctuations of its amplitude quadrature operator are
found to be
\begin{eqnarray}\label{fluctuations}
\delta\hat{X}_{f}^{+}=\sum_{p=0}^{\infty}c_{2p+1}\delta
\hat{X}_{2p+1}^{+},
\end{eqnarray}
where $\delta \hat{X}_{2p+1}^{+}$ corresponds to the fluctuations
of the amplitude quadrature of the mode $u_{2p+1}(x,z)$.

Experimentally, we measure the displacement for different split
detector positions. This displacement is induced by a modulating
device generating at $z=0$ displacement and tilt modulations of
amplitude $d$ and $p$, respectively. A measurement at the
modulation frequency, using a spectrum analyzer yields the
modulation signal as well as the noise at this frequency. As usual
in quantum optics, all equations are directly transposable into
the frequency domain. Using Eq.~\ref{Iminus}, the variance
measured by a spectrum analyzer at the precise modulation
frequency is given by
\begin{widetext}
\begin{center}
\begin{eqnarray}\label{v2SD}
V_{SD}(z)= \kappa NT\left(\frac{\hbar \omega}{ 2\epsilon_{0}c
T}\right)^{2}\left[4\left( NT\right)
\frac{2}{\pi}\left(\frac{d}{w_{0}}\cos{\phi_{G}(z)}+\frac{p
w_{0}}{2}\sin{\phi_{G}(z)}\right)^{2}  +
\langle\left(\sum_{p=0}^{\infty}c_{2p+1}\delta \hat{X}_{2p+1}^{-
(2p+1)\phi_{G}(z)}\right)^{2}\rangle\right],
\end{eqnarray}
\end{center}
\end{widetext}
where $\kappa$ is a constant depending only on the electronic
gains of the spectrum analyzer, $T=1/RBW$ is the integration time
and $c_{1}=\sqrt{2/\pi}$.
The first and second bracketed term in Eq.~\ref{v2SD} respectively
correspond to modulation signal and noise. In the plane of the
modulating device (i.e. for $z=0$), the noise term can be written
$\langle\delta \hat{X}_{f}^{+^{2}}\rangle$ and corresponds to the
noise of the amplitude quadrature of the flipped mode. The flipped
mode is therefore the only mode contributing to the noise in this
particular plane. Note that this is not true all along the
propagation axis. For a coherent incoming beam, this noise term
defines the shot noise level, and is equal to $1$. Note that using
non classical resources for which $\langle\delta
\hat{X}_{f}^{+^{2}}\rangle < 1$ in the detection plane results in
noise reduction. This case will be discussed in section \ref{SQZ}.

The Signal to Noise Ratio (SNR) for a coherent beam is found from
Eq.~\ref{v2SD}
\begin{eqnarray}
SNR_{SD}=4NT
\frac{2}{\pi}\left(\frac{d}{w_{0}}\cos{\phi_{G}(z)}+\frac{p
w_{0}}{2}\sin{\phi_{G}(z)}\right)^{2}. \nonumber
\end{eqnarray}
As stated in section \ref{QNL}, the SNR has a quadratic dependence
in displacement $d$ and momentum $p$.

\subsection{Experiment}

\begin{figure}[!ht]
\begin{center}
\includegraphics[width=8cm]{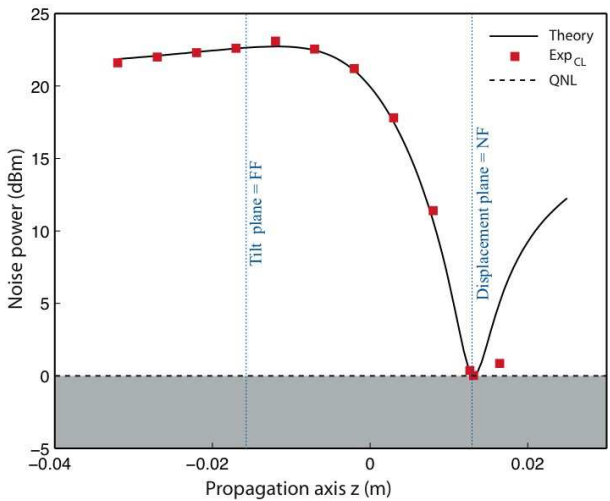}
\caption{{\bf Modulation measurement normalized to the shot noise
along the propagation axis of a tilted and displaced beam, using a
split detector. The modulation was produced by a PZT at 4 MHz, the
near field image is located $1.6~$cm after the waist plane which
is taken as the reference position $z=0$. The modulation detected
in this plane (NF) corresponds to the displacement modulation and
represents only $10~\%$ of the overall modulation strength. The
tilt information lies in its far field (FF).}} \label{SD_exp}
\end{center}
\end{figure}

We have performed split detection measurements of displacement and
tilt of a Gaussian beam, by moving the position of the detector
along the propagation axis, as shown in Fig.~\ref{SD_exp}.
Displacement and tilt are produced by a piezoelectric element
(PZT) modulated at $4~$MHz. Each measurement along the propagation
axis refers to a different quadrature of the modulation (i.e. a
different mixture of displacement and tilt modulation). The
results are normalized to the shot noise and taken with a $4.2~$mW
incident beam. From these measurements, we can infer the
displacement and tilt relative amplitude modulation in the PZT
plane where the waist is $106~\mu$m.  The displacement signal,
accessible in the near field of the PZT, is found to be much
smaller than the tilt signal, and even so that it cannot clearly
be distinguished from the shot noise. This unusual behavior of the
piezoelectric material arises from the operation regime, where the
modulation is generated via an accidental mechanical resonance of
the PZT. The theoretical curve has been plotted for a coherent
illumination, using Eq.~\ref{v2SD} for $10\%$ displacement
modulation, and $90\%$ tilt modulation, ratio determined  with the
more accurate results presented in the section \ref{homodyne}.
There is a very good agreement with the experimental data. The
last experimental point in Fig.~\ref{SD_exp} lies below the
theoretical prediction, as the beam started to be apertured by the
split detector, leading to a smaller measured modulation. Note
that for technical reasons, our experimental setup is slightly
different from the simplest setup presented on
Fig.~\ref{D_and_T_with_SD2}, where the reference plane coincides
with the beam waist position. As shown on Fig.~\ref{SD_exp}, the
waist position lies at $1.6~$cm for the near field of the PZT in
our imaging setup.

We have shown in this section how to retrieve displacement and
tilt information from a gaussian beam with a split detector, and
have taken experimental results which will be used as a reference
in the following sections.

%%%%%%%%%%%%%%%%%%%%%%%%%%%%%%%%%%%%%%%%%%%%%%%%%%%
\section{Homodyne detection with a TEM$_{10}$ local oscillator}\label{homodyne}

We have proved theoretically in reference \cite{Hsu-disp} that
split detection was non optimal to retrieve displacement
information, as it is only sensitive to the flipped mode instead
of the TEM$_{10}$ mode component of the input field. In order to
extract all the displacement and tilt information with up to $100
\%$ efficiency, we propose a homodyne detector involving a
TEM$_{10}$ mode local oscillator, which selects the TEM$_{10}$
mode component of the field.

In the homodyne detection scheme, two beams are mixed on a $50:50$
beam-splitter. The first one is the signal beam containing the
displacement and tilt modulations, whose field operator is given
by Eq.~\ref{field}. The second one is the local oscillator (LO),
whose field operator is
\begin{eqnarray}\label{LO}
\hat{\mathcal{E}}^{+}_{LO}(x) = i \sqrt{\frac{\hbar \omega}{2
\epsilon_{0} c T}}\left[ \sqrt{N_{LO}}u_{1}(x)+
\sum_{i=0}^{\infty}\delta\hat{a}_{LO_{n}}u_{n}(x)\right],\nonumber
\end{eqnarray}
where $N_{LO}$ denotes the number of photons per second in the
beam, and $\phi_{LO}$ is the local oscillator phase. Since
displacement and tilt modulations are very small and the local
oscillator is much brighter than the signal beam (i.e. $N_{LO}\gg
N$), the usual calculation of the intensity difference between the
two homodyne detectors at the modulation frequency gives
\begin{eqnarray}\label{HD_intensity_difference}
\hat{I}_{-} &=& \frac{\hbar \omega}{2 \epsilon_{0} c T}\left[
2\sqrt{NN_{LO}}\left(\frac{d}{w_{0}}\cos{\phi_{LO}}+ \frac{w_{0}p }{2}\sin{\phi_{LO}}\right)\right.\nonumber \\
&+&\left.\sqrt{N_{LO}}\delta\hat{X}_{1}^{\phi_{LO}}\right],
\end{eqnarray}
and similarly to Eq.~\ref{v2SD}, its variance at the displacement
and tilt modulation frequency is therefore
\begin{widetext}
\begin{center}
\begin{eqnarray}\label{v2HD}
V_{HD}(\phi_{LO})  =  \kappa N_{LO}T \left(\frac{\hbar \omega}{2
\epsilon_{0}c T}\right)^{2}\left[ 4NT \left(
cos(\phi_{LO})\frac{d}{w_{0}}+ sin(\phi_{LO})\frac{p
w_{0}}{2}\right)^{2} +\langle\delta \hat{X}_{1}^{\phi_{LO}^{
2}}\rangle \right],
\end{eqnarray}
\end{center}
\end{widetext}
where the constant $\kappa$ is identical to the split detection
part as long as the spectrum analyzer settings have not been
changed. The first bracketed term corresponds to the modulation
signal. The second one refers to the noise of the TEM$_{10}$
component of the detected field, and its variation with the local
oscillator phase $\phi_{LO}$ is given by $\langle\delta
\hat{X}_{1}^{\phi_{LO}^{ 2}}\rangle=\langle\delta \hat{X}_{1}^{+^{
2}}\rangle \cos^{2}{\phi_{LO}} + \langle\delta \hat{X}_{1}^{- ^{
2}}\rangle \sin^{2}{\phi_{LO}}$, where $\langle\delta
\hat{X}_{1}^{+^{ 2}}\rangle$ and $\langle\delta \hat{X}_{1}^{-^{
2}}\rangle$ are the noise of the amplitude and phase quadrature of
the TEM$_{10}$ mode, respectively. Scanning the local oscillator
phase allows to measure all the quadratures of the displacement
and tilt modulation. We have omitted the Gouy phase shift in the
previous expression, as it can be incorporated as a constant term
in the local oscillator phase. This phase is still defined so that
$ \phi_{LO}=0$ corresponds to a displacement measurement in the
PZT plane.

Only the TEM$_{10}$ mode of the incoming beam contributes to the
noise, as it matches the local oscillator transverse shape. All
the other modes contributions cancel out since they are orthogonal
to the local oscillator. The TEM$_{10}$ mode is thus the noise
mode of the homodyne detection, and precisely matches the
information to be extracted. We can show, using a Cramer Rao bound
estimation, that the TEM$_{10}$ homodyne detection is an optimal
displacement and tilt detection, as no other device can possibly
perform such measurements with a better SNR \cite{CRB}.

For a coherent incoming TEM$_{10}$ mode, the previous noise term
defines the shot noise level, and is equal to $1$. Using squeezed
light in the TEM$_{10}$ mode component of the incoming beam would
result in noise reduction, and will be discussed in the section
\ref{SQZ}.

The SNR for a coherent incoming beam can be derived from
Eq.~\ref{v2HD} in the homodyne detection case
\begin{eqnarray}
SNR_{HD}=4NT\left(\frac{d}{w_{0}}\cos{\phi_{LO}}+\frac{p
w_{0}}{2}\sin{\phi_{LO}}\right)^{2}.
\end{eqnarray}

Comparing the split and homodyne detections schemes yields certain
similarities between Eq.(\ref{v2SD}) and (\ref{v2HD}). First, a
variation of the local oscillator phase $\phi_{LO}$ in the
homodyne scheme is equivalent to a propagation along the $z$ axis
inducing a Gouy phase shift $\phi_{G}$ in the split detection
case. Secondly, an additional $2/\pi$ geometry factor in the split
detection case arises from the imperfect overlap between the
flipped mode and the TEM$_{10}$ mode, as discussed in reference
\cite{Hsu-disp}. The comparison between the two SNRs in the
coherent case yields a theoretical efficiency ratio given by
\begin{eqnarray}\label{efficiency}
R_{th}=\frac{SNR_{SD}}{SNR_{HD}}=\frac{2}{\pi}\frac{N_{SD}}{N_{HD}},
\end{eqnarray}
where $N_{SD}$ and $N_{HD}$ refer to the number of photon per
second in the displaced and tilted beam, for the split detection
and the homodyne detection case, respectively. For identical
signal beams powers, this means that the split detection is only
$2/\pi=64\%$ efficient compared to the homodyne detection. Using
the homodyne detection thus corresponds to an improvement of
$(100-64)/64=56\%$.

Eventually, the intensity factor before the bracketed term in
Eq.~\ref{v2HD} and Eq.~\ref{v2SD} can be much bigger in the
homodyne detection case, as it corresponds to the local oscillator
intensity instead of the input beam intensity in the split
detection case. It is thus easier to have more electronic noise
clearance in the homodyne case.

In this section, we have shown - still theoretically - how to
retrieve displacement and tilt using a homodyne detector with a
TEM$_{10}$ local oscillator. Moreover, we have proved a $56\%$
theoretical improvement of this scheme compared to the split
detection.
%%%%%%%%%%%%%%%%%%%%%%%%%%%%%%%%%%%%%%%%%%%

\section{Displacement and Tilt Measurements beyond the Quantum Noise Limit}\label{SQZ}

When the information to be retrieved is below - or of the order of
- the quantum noise, non classical resources (i.e. squeezed laser
beams) can help extracting the information. For each type of
detection (i.e homodyne- and split detection), the only transverse
mode component within the incident field which contributes to the
noise has been identified in the previous sections. The noise
modes of the split and homodyne detection are the flipped mode and
the TEM$_{10}$ mode, respectively. Since displacement and tilt of
a TEM$_{00}$ beam lies in the TEM$_{10}$ component of the beam,
noise mode and information encoded are matched for the homodyne
detection only, accounting for the non optimal split detection.

\begin{figure}[htbp]
\begin{center}
\includegraphics[width=8cm]{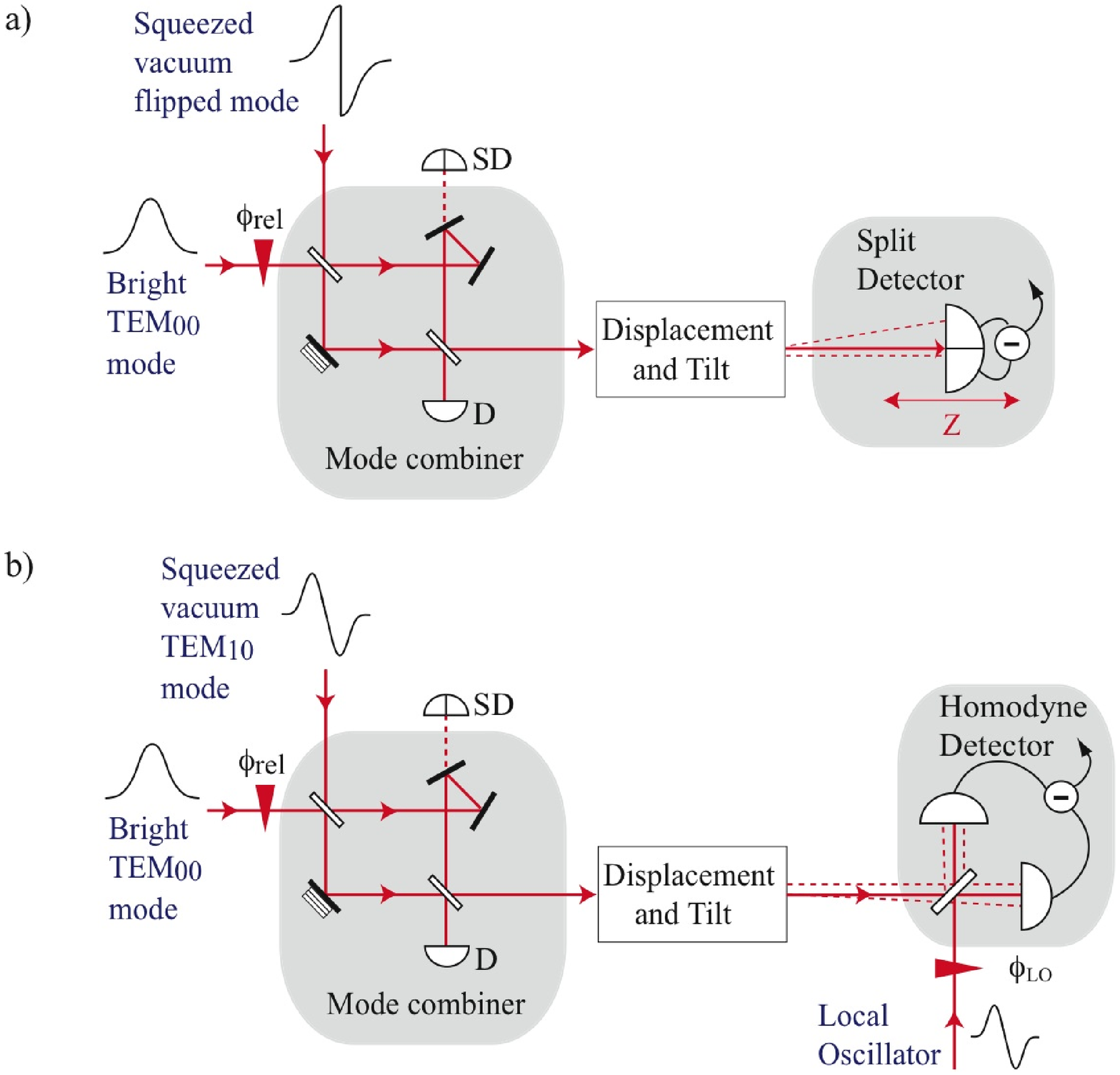}
\caption{{\bf Schematic of displacement and tilt modulation
measurement beyond the quantum noise limit. a) With a split
detector and b) with a homodyne detector. Prior to the modulation
generated via a PZT at a few MHz, a bright TEM$_{00}$ beam is
combined without losses with a squeezed vacuum noise mode. This
was done with a special Mach-Zehnder which has an additional
mirror in one arm. A mirror leakage is used to lock the relative
phase between both input modes. All different combinations of
displacement and tilt modulations are accessible when a) the
position of the split detector along the propagation axis $z$ is
varied, and b) when the local oscillator phase $\phi_{LO}$ is
scanned.}} \label{SD_and_HD_scheme}
\end{center}
\end{figure}

Sub shot noise measurements with both schemes can be performed
using the setups shown in Fig.~\ref{SD_and_HD_scheme}, by filling
the noise mode of the input beam with squeezed light. A mode
combiner has to be used to merge the signal beam - in our case a
bright TEM$_{00}$ beam - with the noise mode of detection, filled
with squeezed vacuum. Note that it has to be a vacuum mode, or a
very dim field not to contribute to the signal, but only to reduce
the quantum noise properties. The combination of beams cannot be
done with a sheer beam-splitter as the squeezing is not robust to
losses. Instead, we used a special Mach-Zehnder interferometer
with an additional mirror in one arm, see
(Fig.~\ref{SD_and_HD_scheme}). This mirror has no effect on even
transverse profiles, but induces an additional $\pi$ phase shift
to odd transverse profiles. Therefore, thanks to this asymmetry,
orthogonal even and odd modes, which are incident on the two input
ports of the Mach-Zehnder, interfere constructively on the same
output port without experiencing any losses. The integrality of
the bright beam and the squeezing of the squeezed vacuum mode - a)
flipped mode or b) TEM$_{10}$ mode - are thus preserved at the
output of the interferometer. Note that other devices can be used
\cite{treps1d, treps2d, treps2} .
%
%Treps {\it et al.} have recently demonstrated nano-displacement
%measurements of optical beams using a split-detector
%\cite{treps1d, treps2d, treps2}. They used a ring cavity for the
%combination of beams. The cavity was locked to resonant for the
%TEM$_{00}$ mode - which was therefore transmitted - but not for
%the flipped mode - which was therefore reflected. In this way, the
%two orthogonal modes were combined without losses
%\cite{treps2,treps2d}. Their measurements were limited to
%displacement in the split detector plane, and were not therefore
%fully characterizing the modulation created onto the beam.
%Moreover, although their scheme was optimized for split detection,
%the theoretical efficiency of such a measurement is limited to
%$\sim~64\%$ for a coherent beam as just discussed above.

\begin{figure}[htbp]
\begin{center}
\includegraphics[width=8cm]{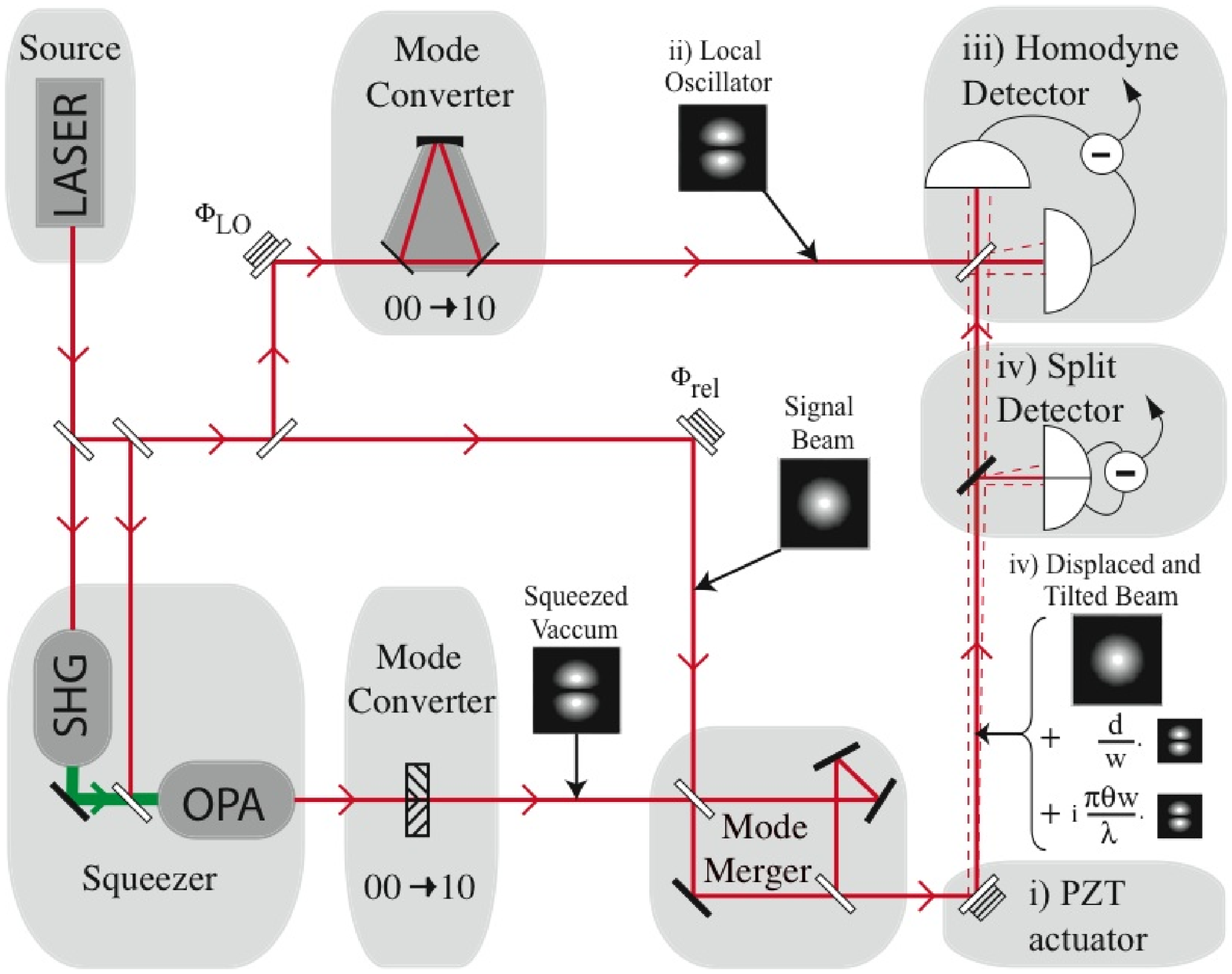}
\caption{{\bf Experimental scheme to measure displacement and tilt
with a split detector and a homodyne detector with the same
operating conditions.}} \label{global_scheme}
\end{center}
\end{figure}

In order to make a direct comparison of the performances of the
split detection and the homodyne detection, we have built the
experimental setup sketched in Fig.~\ref{global_scheme}, where
both schemes are tested in the same operating conditions. In
addition to a simple comparison involving only classical
resources, we designed the experience in order to allow
measurements beyond the QNL. At this stage, we were unable to
produce directly a squeezed TEM$_{10}$ mode, we have therefore
chosen to generate a squeezed flipped mode, which also corresponds
to a squeezed TEM$_{10}$ mode having experienced $36~\%$ losses.
Indeed, the amount of squeezing in the amplitude quadrature of the
TEM$_{10}$ component of the flipped mode can be deduced from
\begin{eqnarray}
\langle\delta \hat{X}^{+
^{2}}_{1}\rangle=\frac{2}{\pi}\langle\delta \hat{X}^{+ ^{2}
}_{f}\rangle+\left(1-\frac{2}{\pi}\right),
\end{eqnarray}
as all the modes except the flipped mode are filled with coherent
light.
%Note that we can only easily identify the fluctuations of
%each component of the flipped mode for $z=0$, as each mode
%quadrature evolve at different speed during propagation because of
%the Gouy phase shift. The amount of squeezing in the amplitude
%quadrature thus evolves rapidly with propagation. But
If the flipped mode is a classical coherent beam, $\langle\delta
\hat{X}^{+ ^{2}}_{f}\rangle=1$, which also implies that the
TEM$_{10}$ component is coherent $\langle\delta \hat{X}^{+^{
2}}_{1}\rangle=1$, as expected. In the end, if we start with
$3.6~$dB of squeezing in the amplitude quadrature of the flipped
mode as discussed below, we get $2~$dB of squeezing in the
TEM$_{10}$ mode, which is exactly what can be measured
experimentally by using the homodyne detection with a TEM$_{10}$
local oscillator.

We used the following experimental procedure. First we generated a
$3.6~$dB squeezed TEM$_{00}$ mode from a monolithic optical
parametric amplifier (OPA) pumped by a frequency doubled YAG laser
delivering $600~$mW at $1064~$nm, and seeded by a TEM$_{00}$ mode.
This very low power (nW) squeezed beam then experiences a mode
conversion into the flipped mode thanks to a special wave-plate
made of two half wave-plates whose optical axis have been rotated
$90 ^{\circ}$ relative to each other \cite{Delaubert-flipped}. A
beam incident on such an optical element yields a $\pi$ phase
shift on half of its transverse profile.

Thanks to the special Mach-Zehnder interferometer formerly
presented, we combine this beam with a bright TEM$_{00}$ beam,
therefore preserving their potential non classical properties. To
achieve this experimentally, we first mode matched both input
beams of the interferometer without the special wave-plate,
reaching $99.5~\%$ visibility on the first beam-splitter of the
interferometer. The squeezed beam, although very dim, is still
bright enough to be mode matched with the other bright TEM$_{00}$
beam. The interferometer is then aligned on the OPA beam without
the wave-plate with $98~\%$ visibility and then the wave-plate is
slid in the center of the beam to a maximum visibility of
$97.8~\%$. Note that we purposely introduced a leakage in one of
the mirrors to lock the relative phase $\phi_{rel}$ between the
two input modes with a split detector (SD), as drawn in Fig.~
\ref{SD_and_HD_scheme}). In the end, the global mode combiner
efficiency is still higher than $97\%$.

The multi-mode squeezed beam hereby generated is then displaced
and tilted with a PZT, as presented in section \ref{split}, and
the information is detected with either one of the split or
homodyne detection schemes. We only presented classical
measurements for the split detection case in Fig.~\ref{SD_exp}, as
such measurements have already been demonstrated. Moreover, the
measurements have to be done precisely in the near field of the
mode converting wave-plate - which is also the near field of the
PZT - as the flipped mode is not stable in propagation, and
squeezing degrades very quickly along the $z$ direction.

The TEM$_{10}$ local oscillator is produced with a misaligned ring
cavity locked to resonance on the TEM$_{10}$ mode represented in
Fig.~\ref{global_scheme}. The cavity has been designed such that
it delivers a pure transverse output mode (i.e. high order modes
are not simultaneously resonant in the cavity). We mode matched
this local oscillator beam to the signal beam by previously
locking the ring cavity to the TEM$_{00}$ mode resonance, reaching
a visibility of $98.9~\%$ with the TEM$_{00}$ input mode.

\begin{figure}[h!]
\begin{center}
\includegraphics[width=8cm]{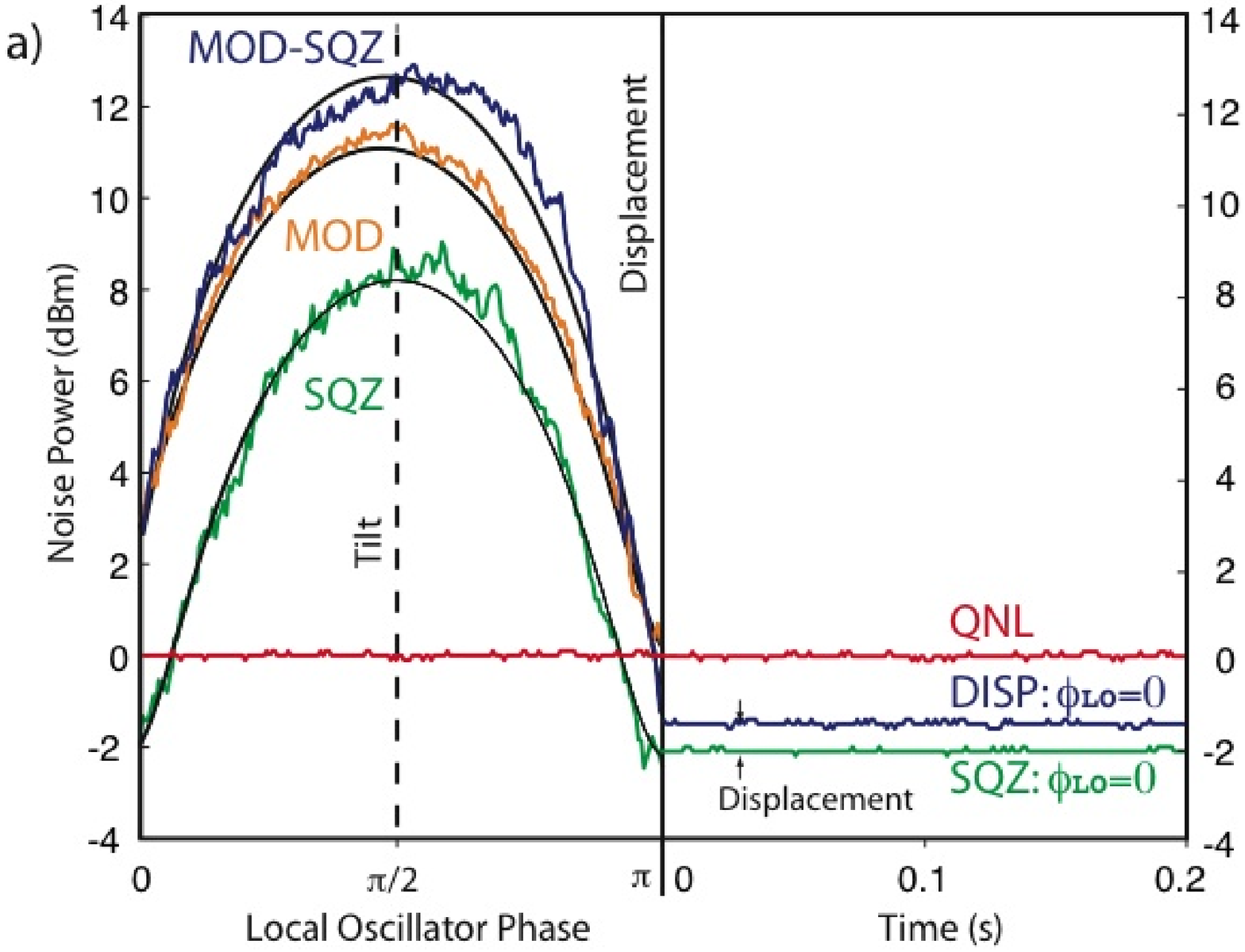}
\includegraphics[width=8cm]{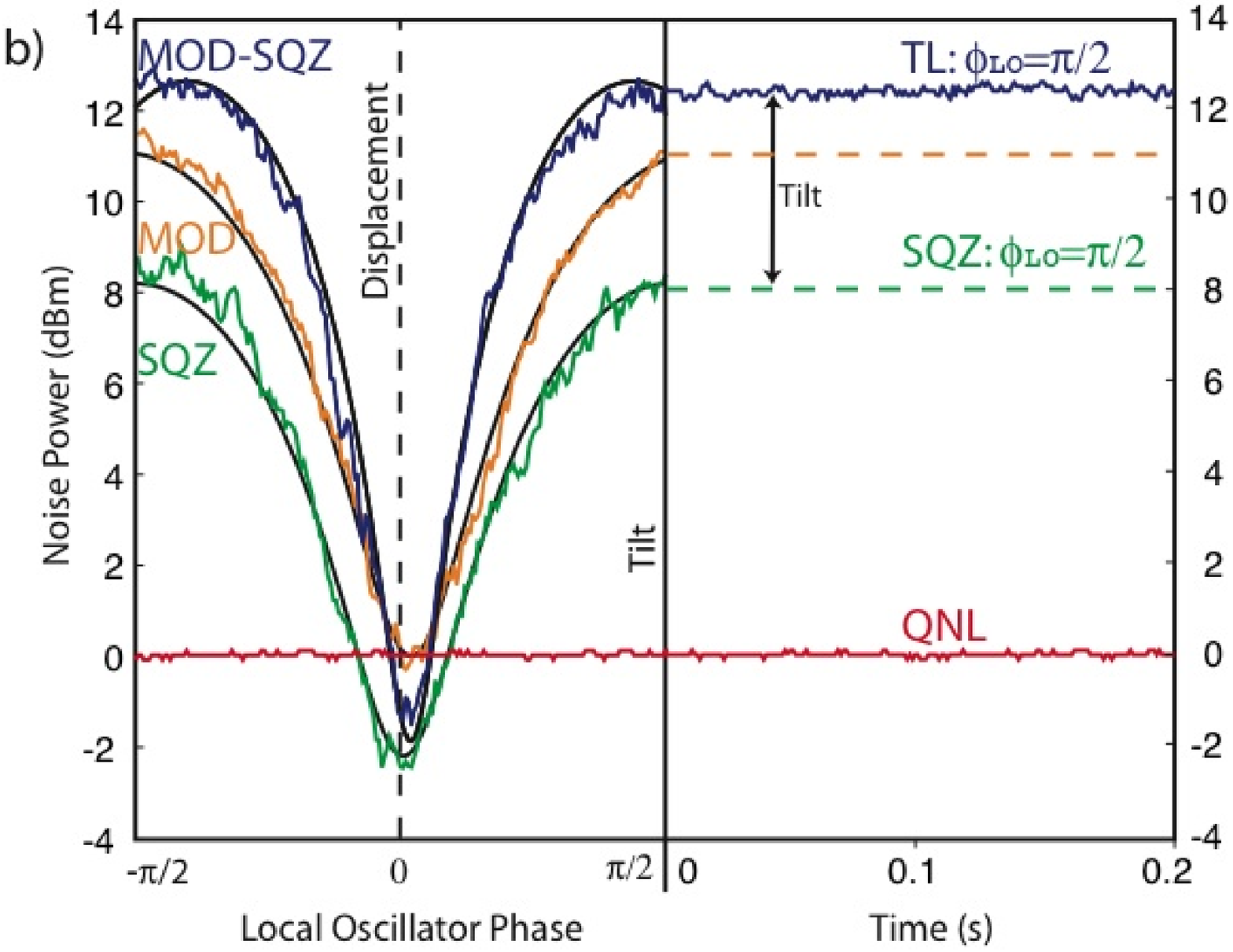}
\caption{Demonstration of sub-shot noise measurements of (a)
displacement and (b) tilt modulations using spatial homodyne
detector.  The figures show an example with $90\%$ of tilt, and
$10\%$ of displacement modulations.  Left hand side of the figures
shows the scanning of the local oscillator phase $\phi_{\rm LO}$
that continuously access the pure displacement (at $\phi_{\rm LO}
= 0 $ and $\pi$) to pure tilt (at $\phi_{\rm LO} = \pi/2$ and
$3\pi/2$) information of the beam.  QNL: quantum noise limit. SQZ:
quadrature noise of squeezed light with 2~dB of squeezing and 8~dB
of anti-squeezing on the TEM$_{10}$ mode, but without any
modulation signal.  MOD: measured modulation with coherent light.
MOD-SQZ: measured modulation with squeezed light. Right hand side
of the figures shows the corresponding locked local oscillator
phase to the (a) displacement or (b) tilt measurement. SQZ: at
$\phi_{LO}=0$ the squeezed noise level is 2~dB below the shot
noise and at $\phi_{LO}=\pi / 2$ there is 8~dB of anti-squeezing
noise.  DISP: MOD-SQZ curve locked to $\phi_{LO}=0$ for
displacement measurement. TL: MOD-SQZ curve locked to
$\phi_{LO}=\pi / 2$ for tilt measurement.  Displacement
measurement is improved by the 2~dB of squeezing, while the tilt
measurement is degraded by the 8~dB of anti-squeezing.}
\label{results}
\end{center}
\end{figure}

The experimental results, obtained with the spectrum analyzer in
zero-span mode at 4~MHz, are presented in Fig.~\ref{results}(a)
and Fig.~\ref{results}(b), when the TEM$_{10}$ local oscillator
phase is scanned and locked for displacement ($\phi_{LO}=0$) and
tilt ($\phi_{LO}=\pi / 2$) measurement. Without the use of
squeezed light, the displacement modulation cannot clearly be
resolved, as in the split detection case. Improvement of the SNR
for displacement measurement beyond the quantum noise limit is
achieved when the squeezed quadrature of the TEM$_{10}$ mode is in
phase with the displacement measurement quadrature (i.e. in phase
with the incoming TEM$_{00}$ mode). Since we are dealing with
conjugate variables, improving the displacement measurement
degrades the tilt measurement of the same beam, as required by the
anti-squeezing of the other quadrature. The displacement
measurement is improved by the 2~dB of squeezing, whereas the tilt
measurement is degraded by the 8~dB of anti-squeezing. Theoretical
curves calculated with $2~$dB of noise reduction and 8~dB of
anti-squeezing, and $90~\%$ of tilt modulation and $10~\%$ of
displacement modulation - continuous curves in
Fig.~\ref{results}(a) - are in very good agreement with
experimental data.  In our experiment, we have a TEM$_{00}$ waist
size of $w_{0}=106~\mu$m in the PZT plane, a power of $170~\mu$W,
$RBW=100~$kHz and $VBW=100~$Hz, corresponding to a QNL of
$d_{QNL}=0.6~$nm. The measured displacement lies 0.5~dB above the
squeezed noise floor, yielding a displacement modulation 0.08
times larger than the QNL. As the modulation has a square
dependance on the displacement $d$, we get
$d_{exp}=\sqrt{0.08}d_{QNL}=0.15~$nm. This would correspond to a
trace 0.3~dB above the QNL, and cannot therefore be clearly
resolved without squeezed light. The ratio between displacement
and tilt modulations can be inferred from the theoretical fit in
figure \ref{results}, giving a measured tilt of $0.13~\mu$rad.

We have in this section demonstrated measurements of a pair of
quantum conjugate variables - displacement and tilt - with a
homodyne detector involving a TEM$_{10}$ mode local oscillator,
and performed sub shot noise displacement measurements.

%%%%%%%%%%%%%%%%%%%%%%%%%%%%%%%%%%%%%%%%%%%%
\section{Comparison}

We can compare the results obtained with the split detection and
the homodyne detection by using Eq.~\ref{efficiency}, where we
need to take into account the input beam power discrepancy between
both experiments. It was necessary to take data with more power in
the modulated beam for the split detection in order to have a
sufficient noise clearance between the electronic noise and the
shot noise. As for the homodyne detection, the local oscillator
can be intense, and the noise clearance above the electronic noise
can reach around 10~dB. This is interesting when the signal to be
retrieved is very small, but can on the other hand also be a
limitation for large signals. The experimental efficiency ratio
can be evaluated with the following expression
\begin{eqnarray}\label{Refficiency}
R_{exp}=\frac{P_{HD}}{P_{SD}}\left[\frac{Mod_{SD}}{Mod_{HD}}\right]=0.64,
%=\frac{P_{inputHD}}{P_{inputSD}}\left[\frac{10^{\frac{Mod_{SDdB}}{10}}-1}{10^{\frac{Mod_{HDdB}}{10}}-1}\right],
\end{eqnarray}
where $P_{HD}=170 \mu$W and $P_{SD}=4.2~$mW refer to the signal
beam powers, in the case of the homodyne and split detection,
respectively. $Mod_{SD}$ and $Mod_{HD}$ are the maximum detected
modulation in each detection scheme, relative to the shot noise.
The two experimental curves read $Mod_{SD}=23$~dB and
$Mod_{HD}=11.3$~dB, which yields $R_{exp}=0.64$, and which
precisely correspond to the theoretical expectation. The
uncertainty, mainly given by the one on the maximum modulation
values, can be evaluated to $5\%$. We therefore report an
efficiency improvement of $56\%$ in perfect agreement with the
theoretical value calculated in reference \cite{Hsu-disp}.

Although the split detection is easier to set-up experimentally,
the homodyne scheme is more efficient and allows measurements of
both the displacement and tilt of the beam without moving the
detector position, giving a complete information on the
interaction between the sample and the beam. We show that one can
choose which measurement to improve (tilt or displacement) using
squeezed light by simply varying the phase of the squeezed beam.
There are additional limitations to the use of a split detector
due to its gap and to its finite size, which are imposing
constraints when the variation of the modulation on the
propagation axis is measured. The accessible range to a good
detection on the $z$ axis is small, as the beam can neither be too
small (because of the gap), nor too large (because of the finite
size of the detector). Moreover, in order to perform measurements
beyond the QNL, one has to carefully image the squeezed flipped
mode onto the sample, and also onto the detector, as the flipped
mode is not stable in propagation. This imaging difficulty is not
present in the homodyne case. Nevertheless, it is replaced with
the constraint of a careful mode matching of the incoming beams.
Moreover, homodyne detection cannot be used for the positioning of
an incoherent light beam, as it is relying on the interferences
between incoming beam and local oscillator.
\\
\section*{Conclusion}

We have demonstrated a homodyne detection scheme involving a
TEM$_{10}$mode local oscillator in order to measure the
displacement and tilt of a Gaussian beam. We report a detection
efficiency improvement of $56\%$ relative to the split detection,
in perfect agreement with the theoretical value. Our detection
setup is very simple and could thus easily replace split and
quadrant detectors in many applications, particularly when tilt
and displacement are needed at the same time. Moreover, further
developments using non classical are possible, as we have
presented measurements beyond the QNL with these devices. Note
that we are now able to generate the squeezed TEM$_{10}$ mode
directly, without the use of a wave-plate, with a misaligned
Optical Parametric Amplifier \cite{NaturePhys}, allowing a
simplification of the setup.

Quantum measurements in the transverse plane such as the ones
presented in this paper potentially open the way to parallel
quantum information processing. Indeed, instead of using amplitude
and phase quadratures or Stokes operators, conjugated quantum
operators are now available in the transverse plane. The
generation of spatial-entanglement between position and momentum
of two laser beams will be considered as a first step towards this
goal. Note that spatial entanglement has already been demonstrated
with orbital angular momentum in the single photon regime
\cite{Woerdman1, Woerdman2}. Other types of experiments that could
follow are dense coding and teleportation of spatial information
and spatial holography.

\section{Acknowledgements}

We like to thank Magnus Hsu, Warwick Bowen and Nicolai Grosse for
helpful discussions. This work was supported with funding from the
Australian Research Council Centre of Excellence program. P.K.Lam
and H-A.Bachor enjoy funding as ARC fellows.

%%%%%%%%%%%%%%%%%%%%%%%%%%%%%%%%%%%%%%%%%%%

\end{document}